# Design, characterization and visual performance of a new multizone contact lens

Manuel Rodriguez-Vallejo[1*], Clara Llorens-Quintana[3], Juan A. Monsoriu[2] and Walter D. Furlan[3]

[1] Qvision, Unidad de Oftalmología Vithas Hospital Virgen del Mar, 04120, Almería

[2] Centro de Tecnologías Físicas, Universitat Politècnica de València, 46022 Valencia, Spain

[3] Departamento de Óptica, Universitat de València, 46100 Burjassot, Spain

[*]Corresponding author: manuelrodriguezid@qvision.es

## 1. ABSTRACT

**Objectives:** To analyze the whole process involved in the production of a new bifocal Multizone Contact Lens (MCL) for presbyopia.

**Methods:** The optical quality of a new MCL was evaluated by ray tracing software in a model eye with pupil different diameters with the lens centered and decentered. A stock of low addition (+1.5 D) MCL for presbyopia was ordered for manufacturing. Power profiles were measured with a contact lens power mapper, processed with a custom software and compared with the theoretical design. Nine lenses from the stock were fitted to presbyopic subjects and the visual performance was evaluated with new APPs for iPad Retina.

**Results:** Numerical simulations showed that the trough the focus curve provided by MCL has an extended depth of focus. The optical quality was not dependent on pupil size and only decreased for lens decentered with a pupil diameter of 4.5 mm. The manufactured MCL showed a smoothed power profile with a less-defined zones. The bias between experimental and theoretical zone sizes was uniform along the optical zone unless for the most central area. Eyes fitted with the manufactured MCL showed an improvement in near Visual Acuity (VA) and near stereopsis. Althouh Contrast Sensitivity (CS) at distance decreased, the defocus curve for contrast showed an extended depth of focus correlated to the ray tracing results.

**Conclusions:** The understanding of vision with MCL requires a process that involves design and characterization for detecting any defect that may have impact in the final visual performance.

## 2. INTRODUCTION

Presbyopia correction with multifocal contact lenses (MCLs) has been for years one of the most important topics in optometry research from the emergence of early designs on the latter half of the 1980s (Toshida et al. 2008). Two solutions: alternating vision, and simultaneous vision, have been widely studied, being the latter the most popular nowadays (Charman 2014). Simultaneous vision is achieved through varying the power along some areas of the lens in such a way that light is distributed in more than one single focus. This concept has evolved from the design proposed by de Carle (de Carle 1989) with multiple variations including diffractive MCLs and refractive MCLs with centre-distance aspheric, centre-near aspheric or multiple zones (Hough 2006). During the last 20 years, some improvements have been proposed in the design, characterization and visual performance assessment with these MCLs (Plakitsf and Charman 1995). Advanced ray tracing software is currently used to design and simulate the optical performance of the MCLs in model eyes (Bradley et al. 2014; Rodriguez-Vallejo et al. 2014). Manufactured lenses can be precisely characterized by means of new objective instruments (Joannes et al. 2010; Plainis et al. 2013; Wagner et al. 2014), and the visual performance for a wide range of distances can be assessed by through-focus plots (Plainis et al. 2013). The whole process involved in the development of a new design of MCL is a linked chain which must accept feedbacks of the partial results obtained in the process. In this work, we propose an approach that covers all the steps involved in the development of new MCLs, from the optical design to the final visual performance obtained by the observer, through the manufacturing process and the characterization of the prototypes.

## 3. METHODS

*Contact Lenses Modelling*

A new design of bifocal MCL for presbyopia treatment has been designed and evaluated. It consists of 6 refractive zones, (3 for distance correction (D) and 3 for near correction (N)). Zone diameters from center to periphery were 1.76 mm (D), 2.48 mm (N), 3.04 mm (D), 4.30 mm (N), 4.64 (D), and 4.96 mm (N). The theoretical optical performance of the MCL was computed by means of the Through Focus Modulation Transfer Function (TF-MTF) for spatial frequencies of 12, 25 and 50 line pairs per mm (lp/mm) over the Atchison model eye (Atchison 2006), setting the refractive error to emmetropia and taking account the Stiles-Crawford apodisation (Bradley et al. 2014). The characteristics of the model eye fitted with a MCL have been described in detail elsewhere (Rodriguez-Vallejo et al. 2014). In this case, the back vertex power of the correction zones was set to zero in such a way the distance (D) focus was on the retina, while the power of near zones was +1.50 D for producing simultaneously an additional, near (N), focus in front of the retina. The effect on the TF-MTF of decentring the MCL in reference to the pupil center was also computed. Simulations were performed with the ray-tracing software package, Zemax 13 SE (Zemax Development Corporation, Bellevue, WA, USA).

*Manufacturing and Characterization*

MCLs were manufactured with Hioxifilcon A (Benz G5X p-GMA/HEMA) (Benz 2016), which has a 1.401 refractive index (hydrated and at 35º), using a precision lathe (Optoform 40). Two stocks of MCLs from 0.00 D to +2.00 D (in +0.50 D steps) with an addition of +1.50 D were ordered for manufacturing (base curves: 8.4 mm or 8.6 mm; diameter: 14.50 mm). The Nimo TR1504 (LAMBDA-X, Nivelles, Belgium)(Joannes et al. 2010) contact lens power mapper was used to characterize the power profiles of the manufactured lenses. Nimo software (version 4.2.6.0 r477) allows to obtain the power of multizone MCLs but only up to five zones defined by the operator. Therefore for computing the D and N powers of our lenses, we exported the power raw data and we developed a custom function in Matlab (version R2013a, The Math-works, Inc.) for detecting D and N zones in the MCLs and for calculating the mean of power along each zone. To do that, changes in the slope of the power profiles higher than 0.25 D were detected by means of computing the first derivative of the radial power function. These changes represent the maxima and minima at D and N zones respectively and the half of power between a minimum and the consecutive maximum (or vice versa) was assumed as the transition between zones. Then the mean experimental power of each zone was computed.

*Subjects and Visual Performance*

Five presbyopic volunteers with a mean age of 49.8 ± 4.3 years (range 45 – 56 years) participated in the study. Subjects were subjected to a complete eye exam including objective and subjective refraction, and slit-lamp exploration. In order to homogenize the sample, inclusion criteria were: hyperopic patients (0.25 D ≤ Rx ≤1.75 D) who need an addition lower than 1.50 D, with no ocular diseases affecting the visual performance, subjective astigmatism under 0.75 D, and normal binocular function (except one subject who was amblyope and only the healthy eye was included in the sample). The addition was defined as the minimum positive power over the distance subjective refraction to comfortably recognize a high contrast optotype of 20/20 at 40 cm. The research adhered to the tenets of the Declaration of Helsinki, with the research approved by Ethics Commission of University of Valencia, and an informed consent was obtained from all participants.
Subjects were fitted with MCLs from an existing stock previously characterized by the Nimo TR1504 as it has been described in the previous section. Therefore, the power of the MCLs fitted to each eye was the closest one to the spectacle refraction in the stock of MCLs. The best base curve from the both possible values, 8.40 mm or 8.60 mm, was selected by the investigator after evaluating, with a slit lamp, the movement and centration of the soft lens after 15 minutes of wearing. Then, the visual performance measurement was conducted with the selected lens.
The visual performance of subjects fitted with MCLs was assessed by measuring: visual acuity (VA) (Rodríguez-Vallejo et al. 2016), contrast sensitivity function (CSF) (Rodríguez-Vallejo et al. 2015), stereopsis (ST) (Rodríguez-Vallejo M, Fajardo VC, Furlan WD, Pons A 2014), and through-focus response (TFR) (Fernández et al. 2016) with Applications (APPs) developed for iPad Retina. All subjects were tested monocularly (with the exception of the stereopsis test) first, without compensation, and then, wearing the MCLs. The APPs allow to measure VA and ST at 3 m and also at near at 0.40 m and 0.50 m respectively. CSF and TFR were conducted at 2 m of distance, the latter varying the contrast (in log units) with an optotype of 0.3 logMAR (TFR-C) or varying the visual acuity (logMAR) of a high contrast optotype (TFR-VA). The optotype used for the measurement of TFR was the Snellen E. For TFR-C, the optotype size was 0.3 logMAR which corresponds to a spatial frequency of 50 mm$^{-1}$ or equivalently 15 cpd.(ISO-11979-2 2014) (Holladay et al. 1990). The absolute area under the TFR was calculated considering a baseline of 0.3 logMAR (Wolffsohn et al. 2013) for TFR-VA and -0.6 log units of contrast for TFR-C. Finally, patients were asked about their satisfaction in terms of general quality of vision achieved with the MCL fitted to each individual eye through an ordinal scale ranging from 1 to 5 (very dissatisfied, dissatisfied, neutral, satisfied, and very satisfied).

*Statistical Analysis*

Normal distributions were tested with the Shapiro-Wilk test and non-parametric statistical tests were used for $p < 0.05$. A Bland–Altman procedure was used to assess the agreement between theoretical and experimental zone sizes. Paired t-tests were used for computing mean differences between theoretical MCL power fitted to the eye and experimental MCL power measured with the Nimo for near and far distance zones. The mean differences in the visual performance, with and without MCLs, were also assessed with paired t-tests. The mean and standard deviation of TFR were computed and represented with Matlab and the areas under the TFR curves were calculated with the included *trapz* function. The data were managed using SPSS software version 20 (SPSS Inc., Chicago, IL, USA), and p<0.05 was considered to indicate significance.

## 4. RESULTS

*Contact Lenses Modelling*

Fig. 1 shows the theoretical TF-MTF for the naked eye and pupil diameters of 3.5 mm (panel A) and 4.5 mm (panel D) compared with the same eye with MCLs centered (panels B and E) and decentered (panels C and F). As can be seen, the theoretical effect of the MCL is to produce an extended depth of focus (DOF) with a concomitant myopic focal shift of the whole focal volume. Particularly for the 25 lp/mm frequency, two peaks can be observed: one at 0 mm (infinite in the object space) and at 0.5 mm in front of the retina which is conjugated with a plane at 67 cm in front of the eye. As expected, the best performance of the MCL was obtained for the lens centered on the pupil.

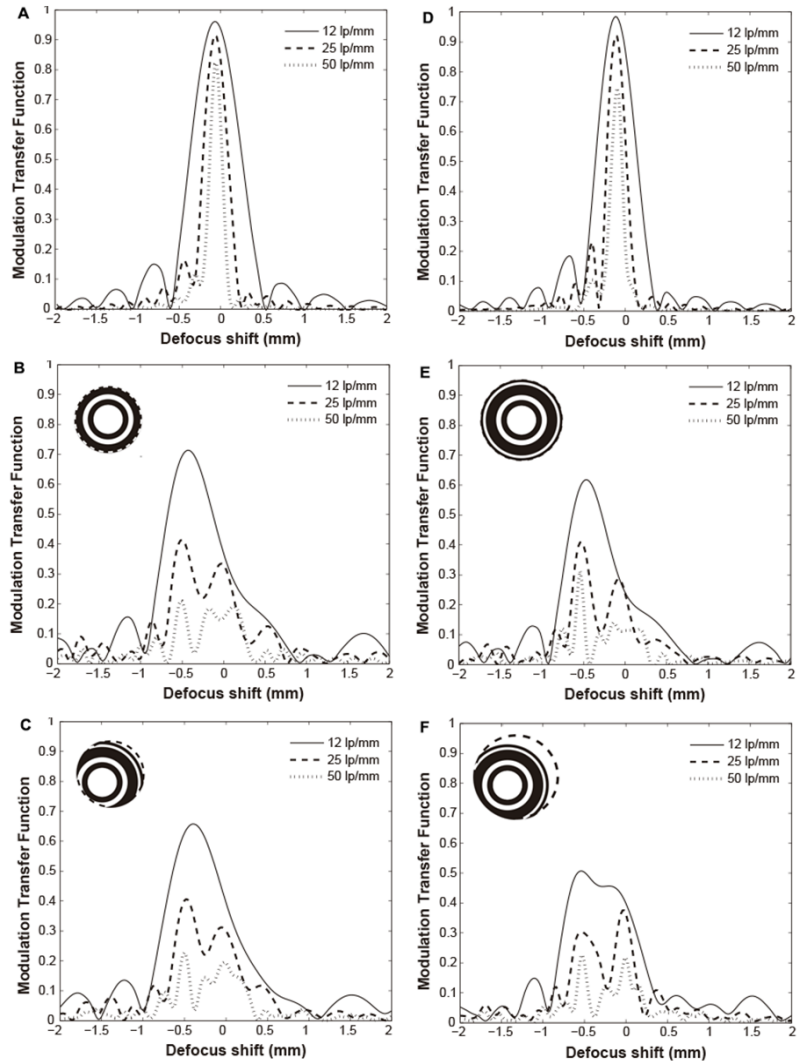

*Figure 1. Theoretical, 12 lp/mm, 25 lp/mm and 50 lp/mm, TF-MTF for an emmetropic model eye. Upper row shows results for pupil diameter of 3.5 mm for the model eye: (A) without CL, (B) with MCL, and (C) with MCL decentered. Bottom row describes results for pupil diameter of 4.5 mm for the model eye: (D) without CL, (E) with MCL, and (F) with MCL decentered. MCL Add was +1.50 D and the decentering was 0.5mm downward and 0.5mm sideward. The inset diagram describes the distance zones (white) and the near zones (black) covering the pupil (dotted circle).*

*Manufacturing and Characterization*

The assessment of the manufacturing process was performed using a customized algorithm as explained in the Methods. A typical result of the power profile of one of the MCLs is shown in Fig. 2. Vertical lines represent the transition zones detected by the algorithm and the horizontal lines along +1.00 D and +2.50 D represent the theoretical powers for distance and near respectively. The grey areas represent the differences between the theoretical power and the experimental power profile. Similar results were found for the nine MCLs fitted to the subjects in the experiment, we found that, compared with theoretical power, the the mean experimental power of the lenses was +0.12 ± 0.22 D for distance [$t(8) = 1.538, p=0.16$] and -0.57 ± 0.19 D for near [$t(8) = -8.9, p < 0.0005$]. Table 1 shows the experimental powers measured at far and near for each MCLs (MCL fitted) and the corresponding theoretical powers of each fitted lens.

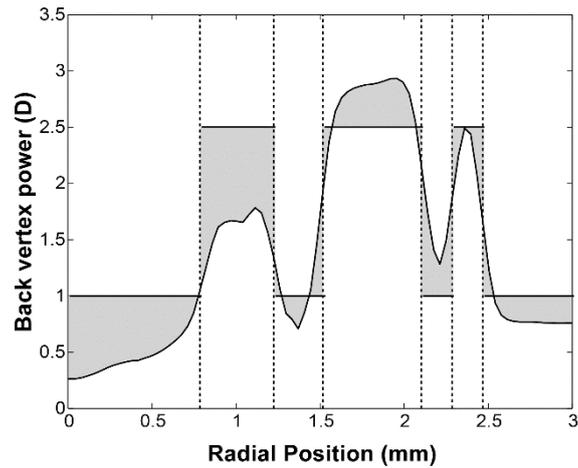

*Figure 2. Power profile of one of the MCLs. The vertical lines represent the transition between zones detected by the algorithm, horizontal lines are the theoretical powers ordered for manufacturing, +1.00 D for distance zones and +2.50 D for near zones in this case. Grey areas show the difference between the experimental power and the theoretical power.*

*Table 1. Spectacle Rx and powers for MCL fitted to each eye in the sample (Experimental power). The theoretical powers of each lens are also included.*

| Subject Id | Eye | Spectacle Rx Far/Near | Experimental Power Far/Near | Theoretical Power Far/Near |
|---|---|---|---|---|
| 1 | RE | +0.25 /+1.25 | +0.61 /+0.70 | +0.00 /+1.50 |
|   | LE | +0.50 /+1.50 | +0.54 /+1.39 | +0.50 /+2.00 |
| 2 | RE | +1.00 /+2.00 | +0.30 /+1.72 | +0.50 /+2.00 |
|   | LE | +1.50 /+2.50 | +1.55 /+2.39 | +1.50 /+3.00 |
| 3 | RE | +1.00 /+2.00 | +1.03 /+2.09 | +1.00 /+2.50 |
|   | LE | +0.50 /+1.50 | +0.54 /+1.39 | +0.50 /+2.00 |
| 4 | RE | +1.00 /+2.00 | +1.04 /+2.18 | +1.00 /+2.50 |
|   | LE | +1.25 /+2.25 | +1.74 /+2.22 | +1.50 /+3.00 |
| 5 | RE | +1.75 /+3.50 | +2.17 /+2.78 | +2.00 /+3.50 |
| mean |  | +0.97 /+2.06 | +1.06 /+1.87 | +0.94 /+2.44 |
| ±SD |  | +0.49 /+0.67 | +0.64 /+0.63 | +0.63 /+0.63 |

In general, we found that the bias between theoretical (*Pt*) and experimental (*Pe*) powers of MCLs was different depending on the zone. Fig. 3A shows lower *Pe* than *Pt* at zones 1 (D) and 2 (N) and the opposite at zones 5 (D) and 6 (N). A Kruskal-Wallis H test was conducted to determine if there were differences in the power bias between zones. As can be seen in the box plot the distributions of power bias were not similar for all zones, The distributions of power bias were statistically significantly different between zones, $\chi^2(5)$ = 36.319, p< 0.0005. Pairwise comparisons were performed with a Bonferroni correction for multiple comparisons. This post-hoc analysis revealed statistically significant differences in power bias between central zones (zones 1 and 2) and peripheral zones (zones 5 and 6) (p< 0.05). The agreement between theoretical and experimental diameter of the zones is shown in Fig. 3B. It can be seen that the first zone of the manufactured lenses is smaller than the theoretical one. On the other hand, from zones 2 to 6 the bias between theoretical and experimental zones is reduced except for the 4$^{th}$ zone for which appear a slight overestimation of the experimental diameter.

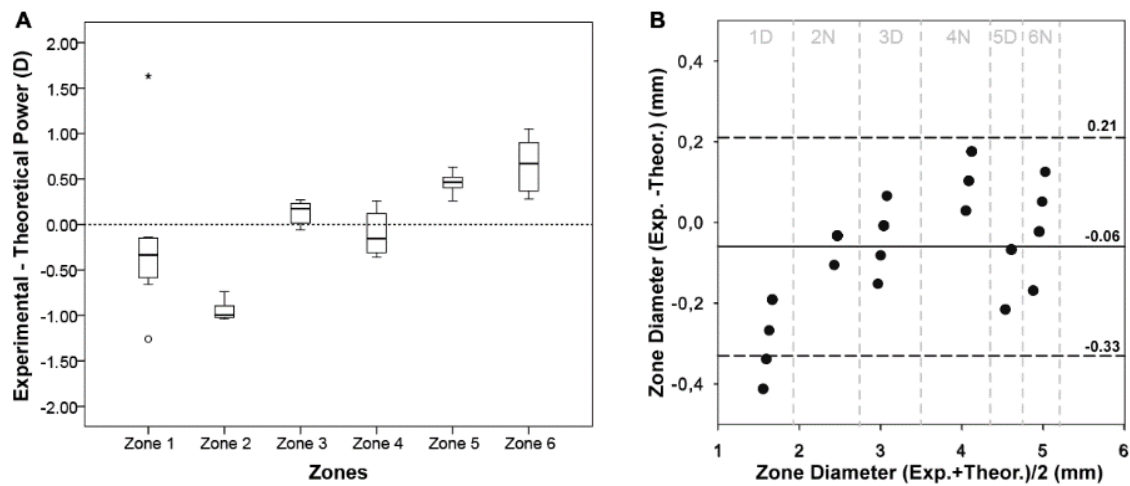

*Figure 3. (A) Difference between the experimental and theoretical powers at each zone (B) Bland-Altman plot. Difference between experimental and theoretical and zone diameters versus mean of experimental and theoretical zone diameters.*

*Clinical Visual Performance*

Table 2 shows that MCLs improved the visual performance at near versus the no presbyopia compensation. Although with the MCLs the improvements ST and VA at near were significant, distance VA was not significantly increased and CSF decreased at frequencies equal or greater than 6 cpd.

Fig. 4 shows the mean TFR performance of the MCLs for the VA (panel A) and C (panel B). The mean area under the TFR-VA and TFR-C curves, was $1.29 \pm 0.32$ and $0.94 \pm 0.56$ respectively. Both TFR curves were positively correlated ($r = 0.741$, p=0.022) exhibiting an extended DOF with a peak at -1.00 D. Mean VA obtained from TFR were $0.1 \pm 0.17$ logMAR at 40 cm, $-0.08 \pm 0.12$ logMAR at 67 cm and $-0.06 \pm 0.1$ logMAR at distance. Mean C were $-0.62 \pm 0.3$ log at 40 cm, $-0.93 \pm 0.23$ log at 67 cm, and $-0.88 \pm 0.19$ log at distance.

*Table 2. Visual performance before and after fitting MCLs for presbyopia correction.*

| Visual performance (distance) | Without FCLs mean ± SD | With FCLs mean ± SD | p-value paired t-test |
|---|---|---|---|
| **Near VA (0.4 m) in logMAR** | 0.46 ± 0.17 | 0.26 ± 0.17 | 0.001 |
| **Far VA (3 m) in logMAR** | 0.09 ± 0.62 | 0.03 ± 0.07 | 0.214 |
| ***Near ST (0.5 m) in arcsec** | 258 [119, 397] | 119 [79, 119] | 0.023 |
| ***Far ST (3 m) in arcsec** | 397 [278, 397] | 397 [40, 397] | 0.157 |
| **CSF (2 m) in log units** | | | |
|  3 cpd | 2.03 ± 0.08 | 1.96 ± 0.16 | 0.084 |
| **6 cpd** | 2.10 ± 0.20 | 1.89 ± 0.22 | 0.044 |
| **12 cpd** | 1.67 ± 0.22 | 1.51 ± 0.12 | 0.045 |
| **18 cpd** | 1.32 ± 0.23 | 1.03 ± 0.14 | 0.005 |

* Median [min, max] and Wilcoxon signed rank test used instead of mean ± SD and paired t-test due to a non-normal distribution of the variable.

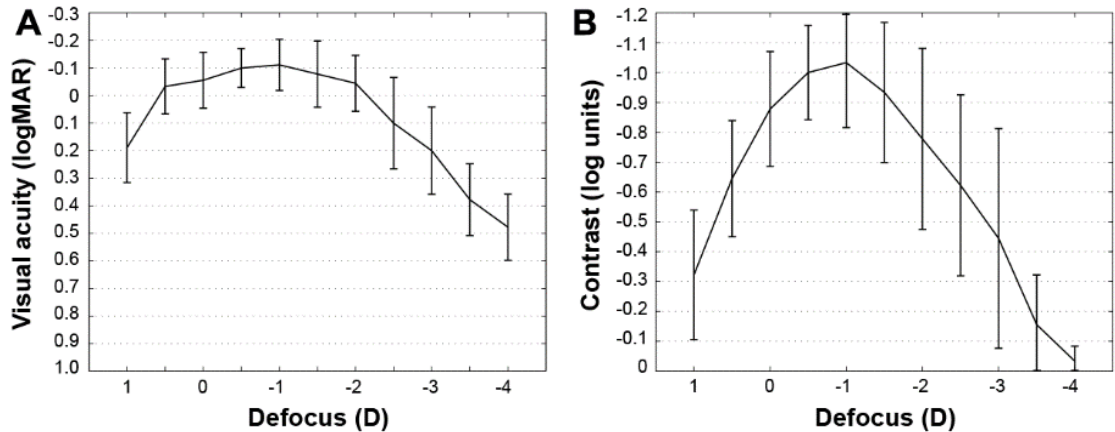

*Figure 4. (A) Through-Focus response for visual acuity (TFR-VA) and (B) for contrast (TFR-C). Vertical bars represent one standard deviation from the mean.*

Satisfaction about the subjective quality of vision achieved with each MCLs was evaluated by means of an ordinal scale, resulting in a mean satisfaction of 3.33 ± 1.33 which corresponded to a value between neutral (3) and satisfied (4). The resulting satisfaction answers were confronted versus the area under the curve for TFR-VA and TFR-C obtained with each lens. The area under the TFR-C was correlated with the general satisfaction r = 0.669 (p = 0.049). No significant correlation (p> 0.05) was found for the area under the TFR-VA.

## 5. DISCUSSION

The optical modelling with Zemax showed an extended DOF performance of the MCL. Interestingly, the bifocal nature of the lens was not expressly revealed for low spatial frequencies but it turned into a myopic focal shift of about 1.0 D or 1.5 D – depending on the pupil diameter. According to our simulations, a loss of contrast in far vision was expected in comparison to the nacked eye, and the TF-MTF predicted a good tolerance to decentrations. The values of decentration were selected according to the approximated values reported with other commercial CLs (Young et al. 2010). The pupil size independence of our design is one of the advantages of the multiple zone CLs in comparison to aspheric CLs as it has been previously reported with other designs (Bradley et al. 2014). However, one potential drawback of the bifocal CLs versus aspheric designs may be a higher incidence of ghost images, mainly for high addition lenses (Kollbaum et al. 2012). In this study this effect might be minimized due to the low addition of our design combined with the smooth transition between zones in the manufactured lenses (Tilia et al. 2016). Further data analysis would be needed in future studies to determine how exactly the modification of our design affects to the ghost images and how the optical simulations are correlated with the real performance in patients fitted with this design.

We used the Nimo TR1504 in order to measure the power profile of our MCL design. In a previous study (Kim et al. 2015), this instrument has demonstrated to be reliable across a half chord between 0.51 mm and 3.2 mm and to have higher variability up to 0.88 D within 1.0 mm diameter. The reliability of this instrument was not tested in this study and we applied our custom algorithm of zones recognition to one measured profile of each lens. We are aware that this could be a limitation of the current work. Future studies should evaluate the repeatability of the construction and assessment procedures. This will help for instance, to elucidate if the bias in the first zone is due to the manufacture or to the repeatability of Nimo and the custom sofware. With respect to this, although power profiles with this instrument have been previously reported by other authors (Kim et al. 2015; Tilia et al. 2016), to our knowledge none have used an algorithm for automatic recognition of zones. Our custom-made algorithm for transition zone recognition has demonstrated to be useful for computing the mean experimental power along each zone and it use could be recommendable instead of using the Nimo TR1504 software. In fact, in this way we were able to detect differences between designed and manufactured lenses either in zone diameters and powers. In general, we found that the lenses showed increasing plus power towards the periphery. Differences in power between corresponding zones could be attributable to a positive spherical aberration of the manufactured plus MCLs. In this sense, testing commercial soft CLs for myopia, Wagner et al. also reported a bias between labeled and measured powers (especially for the central, 1mm, zone) in almost all the CLs measured (Wagner et al. 2014). However, in that work the bias was in the opposite direction than in ours which means that plus CLs might induce positive spherical aberration and minus CLs negative spherical aberration.

The theoretical design of the MCLs of this study corresponds to a multizone refractive CLs. Similar designs are currently available on the market such as the Acuvue Bifocal whose visual performance has been evaluated in multiple studies that reported logMAR VA ranging from -0.11 and 0.19 for distance and from 0.12 to 0.14 for near (Kirschen et al. 1999; Situ et al. 2003; Rajagopalan et al. 2006). Recently, Tilia et al. compared the ACUVUE OASYS for Presbyopia (AOP) and a new extended DOF CL design and they reported mean VAs of -0.06 logMAR and -0.08 logMAR at distance and 0.55 logMAR and 0.40 logMAR at near (40 cm), respectively for low presbyopes (Tilia et al. 2016). In our study, we found similar outcomes of VA as it is reflected in Table 2 and in the VA-TFR. It is important to note that mean VA measured with the APP at near was poorer than the corresponding at -2.5 D of defocus in the TFR APP, 0.26 and 0.1 logMAR respectively. This difference was due to the APP for measuring VA used steps of 0.2 logMAR and the TFR APP used 0.1 logMAR steps. The reason for not using 0.1 logMAR steps with the iPad at 40 cm was the limited of resolution (pixels per inch) of the display (Rodríguez-Vallejo 2016).

In terms of other visual skills, MCLs increased the near stereopsis and decreased the far CSF for spatial frequencies over 6 cpd. The improvement of near stereopsis was related to the improvement of near VA since the random dot stereo-test we used requires non-blurred patterns in both eyes for being correctly fused. Furthermore, the loss in CSF was expected from the comparison between the results shown in Fig. 1 for the nacked-eye and for eye fitted with the MCL.

Another interesting finding is that, even though Zemax simulations predicted a bifocal behavior of our MCLs, experimental TFRs showed smooth extended DOF. This may be due to the non-abrupt transition between near and far zones in the manufactured lenses and to the interaction between the high order aberrations of the eye and the MCLs design (Martin and Roorda 2003).

The general quality of vision achieved with the MCLs abtained a satisfaction grade between neutral and satisfied. The correlations between area under the TFR-VA and TFR-C were computed in order to elucidate if a change in this area might be used in order to predict the satisfaction of the patient depending on the visual performance achieved with the lens. Area under the TFR-VA was not correlated with the satisfaction of the subjects.

However, a correlation was found for the area under the TFR-C. This means that, rather than the TFR-VA, the TFR-C could be a metric which might better predict the satisfaction of the patient. This metric might also be useful to evaluate the agreement between the theoretical optical performance and the vision achieved by the patients and to perform changes in the design based in this agreement. These hypothesis should be confirmed in future studies with a higher sample of subjects.

In this study, we have shown an approach that covers all the steps involved in the development of new MCLs. The whole process from the design to the final MCLs fitted to the patient should include appropriate methods for lens characterization. This will ensure that the reproducibility of the manufacturing process is inside the tolerance limits. Furthermore, in this study new iPad applications have been used satisfactory for measuring the visual performance of subjects wearing MCLs.

## 6. ACKNOWLEDGEMENTS

This study was supported by the Ministerio de Economía y Competitividad (Grant DPI2015-71256-R) and by the Generalitat Valenciana (Grant PROMETEOII-2014-072), Spain.

## 7. CONFLICT OF INTEREST

WDF and JAM have proprietary interest in the contact lens design described in this article. MR-V has designed and programmed the APPs used in this study, which currently distributes by the Apple Store with his own developer account.